\documentclass[twocolumn,showpacs,pre,nofootinbib]{revtex4}
\usepackage{graphicx}
\usepackage{amsmath}
\usepackage{amssymb}
\usepackage{amscd}
\usepackage[dvips]{epsfig}

\newcommand{\clT}{{\cal T}}
\newcommand{\clL}{{\cal L}}

\newcommand{\clR}{{\cal R}}
\newcommand{\bfR}{{\bf R}}
\newcommand{\bfV}{{\bf V}}
\newcommand{\hM}{\hat{M}}

\newcommand{\rgl}{\rangle}
\newcommand{\lgl}{\langle}

\newcommand{\vep}{\varepsilon}

\newcommand{\hh}{\hbar_{\rm eff}}

\begin{document}
\title{From power law to Anderson localization in \\
nonlinear Schr\"odinger equation  with nonlinear randomness}

\author{Alexander Iomin}

\affiliation{Department of Physics,
Technion, Haifa, 32000, Israel.}

\published{: PHYSICAL REVIEW E \textbf{100}, 052123 (2019)}
\begin{abstract}
We study the propagation of coherent waves in a
nonlinearly-induced random potential, and find
regimes of self-organized criticality and other
regimes where the nonlinear equivalent of Anderson
localization prevails. The regime of self-organized
criticality leads to power-law decay of transport
[Phys. Rev. Lett. 121, 233901 (2018)],
whereas the second regime exhibits exponential decay.
\end{abstract}

\pacs{72.15.Rn, 42.25.Dd, 42.65.k}

\maketitle


\section{Introduction }\label{sec:int}

In this work, we consider a problem of polynomial
to exponential localization in the one-dimension nonlinear
Schr\"odinger equation {NLSE) in a random potential.
A special property of the system is that randomness is incorporated into
nonlinearity in the form $\beta\eta(x)|\psi(x)|^2$, where
$\psi=\psi(x)$ is a wave function, $\eta(x)$ is a random fields
and $\beta$ is a nonlinearity parameter.
This formulation of randomness and nonlinearity differs
essentially from the NLSE in random potential,
which reads \begin{equation}\label{NLSE}
i\partial_t\Psi=-\partial_x^2\Psi+\eta(x)\Psi +\beta|\Psi|^2\Psi\, ,
\end{equation}
The model \eqref{NLSE} has been extensively studied,
and a variety of results has been observed over the years.
In particular, a stationary counterpart of Eq. \eqref{NLSE},
when
\begin{equation}\label{NLSE-a}
\Psi(x,t)=e^{-i\omega t}\psi(x)
\end{equation}
relates to the Anderson localization \cite{And58,LeRa85}
of the stationary states of the NLSE.
Another important task is wave propagation in nonlinear media
\cite{BaKl80,FrSpWa86,DeSo86,RaDo87,DoRa87,GrKi92,IoFi07,FiIoMa08,NgKiRoLi11},
where the problem of spreading of wave packets and
transmission are not simply related
\cite{RaDo87,DoRa87,douRam87,PaScLePa07}, in
contrast with the linear case.
This problem is relevant for experiments in
nonlinear optics, for example disordered photonic lattices
\cite{ScBaFiSe07,Lahini08}, where Anderson localization
was found in presence of nonlinear effects.
This long lasting task is far from being completely solved, and many
fundamental problems are still open in both dynamical and
stationary cases, like Berry phase and the semiclassical
a.k.a. the adiabatical approximation in the NLSE \cite{adiabatic}.

Here, we study the propagation of coherent waves in a
random potential that is induced (nonlinearly)
by the wave itself. The problem was first proposed
in the context of nonlinear optics \cite{SSSES18},
but it is in fact a universal problem, relevant to any
coherent wave system, for example, cold atoms in the
Gross-Pitaevskii regime \cite{GP} and more.
The underlying model is fundamentally different from the
NLSE \eqref{NLSE}, because the random potential is
strictly nonlinear, with a mean value around zero.
Technically, this situation corresponds to study of
stationary solutions of the one-dimensional
NLSE in a random potential, where the latter relates now
to the nonlinear part of the NLSE, and this situation relates to
numerical observation of the power law decay of
diffusive waves \cite{SSSES18} in 1D dielectric
disordered -- nonlinear media.
The model in task reads \cite{SSSES18}
\begin{equation}\label{int-1}
\partial_x^2\psi+\omega\psi +\beta\eta(x)|\psi|^2\psi=0\, ,
\end{equation}
where $\omega$ is the energy of the real stationary solution $\psi$.
Here the variables are chosen in dimensionless units and the Planck
constant is $\hbar=1$. Statistical properties of the random potential
$\eta(x)$ will be specified in what necessarily in the text.
To admit the difference between the NLSE \eqref{NLSE}
(with solution \eqref{NLSE-a}) and Eq. \eqref{int-1},
we name the latter by random nonlinear Schr\"odinger equation (RNLSE).

When $\omega=k^2$, where $k$ is a wave
number\footnote{Without restriction of the generality,
the refractive index of the medium is taken to be one.}
and $\beta/k^2$ is the Kerr coefficient, Eq. \eqref{int-1}
is the Helmholtz equation, which corresponds to the experimental
setup  in Ref. \cite{SSSES18}, where
the power law decay of the intensity of the wave has
been observed numerically. Therefore, polynomial decay of
the wave function is anticipated for the solution of
Eq. \eqref{int-1}.

The paper is organized as follows. In Sec.~\ref{sec:heur}
some heuristic arguments based on the random walk theory
is presented to explain the experimental setup.
Sec.~\ref{sec:dsa} is devoted to the estimation
of the transmission coefficient based on the RNLSE
at the condition of the non-zero constant probability current.
Completely original approach
on Anderson localization is developed in Sec.~\ref{sec:fpe},
and its numerical verification is presented in Sec.~\ref{sec:npt}.
Summary of the results and conclusion are presented
in Sec.~\ref{sec:cocl}.

\section{Heuristic arguments of experimental
setup of Ref. \cite{SSSES18}}\label{sec:heur}

Returning to the experiment on wave diffusion, the power
law decay of the wave transmission has been explain by heuristic
arguments as follows.
Due to the Kerr effect, the transmitting characteristics is a
function of the intensity of the wave $\clT=\clT(I)=\sigma I$.
Therefore, the Boltzman equation, which describes intensity of
the diffusive wave reads
\cite{Ishimaru,Furutsu75}
$dI/dx = -\clT(I) I=-\sigma I^2$, with the boundary condition
$I(x=0)=I_0$. This equation defines the power law decay of the
propagating wave amplitude: $I(L)=I_0/(1+\clT(I_0)L)$.
It is a simplified scheme of a more sophisticated toy model
suggested in Ref. \cite{SSSES18}.

\subsection{Random walk approximation}

Kinetic theory of diffusive light in the slab geometry can be
also considered in the framework of random walk theory based on the
universality of the probability of escape from a half space
\cite{Frish95,DaMa97}. This phenomenology is completely relevant
for the multi-collision dynamics for transmission
through a finite slab, considering diffusive waves as a Brownian
particle. In this approach, the transmission probability is determined
by a first passage of a Brownian particle at $x=L$.

A random walk of a particle in random media,
starting at $x(t=0)=0$,
after $n$ identically distributed steps $\Delta x(t_j)$, related to
$n-1$ collisions, is finishing at a random position
$x(t_n)=\sum_{j=0}^{n-1}\Delta x(t_j)$, with the mean squared displacement
$\lgl (x(t_n))^2\rgl \sim \sigma^2 t_n$. Note that the free
pass variance  between collisions $\sigma^2=\lgl (x(t_n))^2\rgl$
is a well defined value
from the experimental setup. Therefore, the mean transition time reads
$t_L\sim (L/\sigma)^2$ for $\sigma\ll L<\infty$ \cite{DaMa97}.
It has been shown in Ref. \cite{Frish95} that the probability of reaching
the boundary $L$, which corresponds to transmission and is determined
as the superposition of all first boundary passages,
reads\footnote{Follow the Sparre Andersen theorem \cite{SparreAnd53},
one obtains that the first passage
probability to escape from a half infinite line for any symmetrical
random walk reads $P_{\infty}(t)\sim t^{-3/2}$.
Therefore, to find a particle outside the boundary $L$
after the mean transit time is
${\rm Prob}(t>t_L)=\int_{t_L}^{\infty}P_{\infty}(t)\sim t_L^{-1/2}$
for the asymptotically large $L$.}
\begin{equation}\label{heur-1}
\clT_L\sim {\rm Prob}(t>t_L)\sim
\frac{1}{\sqrt{\pi t_L}}\sim (L/\sigma)^{-1}\, .
\end{equation}

Although these heuristic arguments on wave diffusion,
based on the either Boltzman equation or random walks,
provide physically reasonable and relevant explanations,
these approaches are far from analytical rigor, related to Eq.
\eqref{int-1}. Another fundamental question is about a localization
length, or Lyapunov exponents of the stationary solution of the RNLSE
\eqref{int-1}. As it has been shown for the
NLSE \eqref{NLSE} in Ref. \cite{IoFi07}, the nonlinearity parameter $\beta$
does not contribute to the Lyapunov exponents of the linear counterpart.
It is also well known that in the linear case for a random
system of the finite length $L$, the transmission
coefficient decays exponentially
with $L$, including the linear part of the NLSE \eqref{NLSE}
\cite{DeSo86,IoFi07}. However, a specific feature of RNLSE \eqref{int-1}
is that for $\beta=0$, the medium is transparent and
the transmission coefficient does not decay at all.
Therefore, our next consideration of the transmission is
in the framework of Eq. \eqref{int-1}.

\section{RNLSE as a Helmholtz equation:
Devillard - Soulliard approach} \label{sec:dsa}

In this section, we investigate the initial-value problem,
where the wave is launched from $x=0$ with some initial
amplitude and induces the nonlinear changes in the
potential as it propagates. In this regime, we
find that the wave follows self-organized criticality,
as it exhibits power law decay while propagating into the structure.

Let us consider one dimensional
wave propagation in the slab geometry, which is
described by Eq. \eqref{int-1}
\begin{equation}\label{dsa-1}
\partial_x^2\psi+\omega\psi +\beta\eta(x)|\psi|^2\psi=0\, .
\end{equation}
The boundary conditions for the random potential are
$\eta(x)=0$ for $x<0$ and $x>L$. Therefore the incident
and reflected (with coefficient $\clR$) waves on the left
and outgoing (with transmission coefficient $\clT$) wave at the
right read
\begin{subequations}\label{dsa-2}
\begin{align}
&\psi(x)=e^{ikx}+\clR e^{-ikx}\, , \quad\quad &x< 0\, , \label{dsa-2a} \\
&\psi(x)=\clT e^{ikx}\, , \quad\quad &x>L\, . \label{dsa-2b}
\end{align}
\end{subequations}
For the fixed output condition, the conservation condition
for the current reads
\begin{equation}\label{dsa-3}
J(x)=[\psi^*\partial\psi-\psi\partial\psi^*]/2i=|\clT|^2=1-|\clR|^2\, .
\end{equation}

In this section, we follow Devillard and Souillard consideration
in Ref. \cite{DeSo86}. Namely, we follow their improved
(theoretically and numerically) Theorem (3), which states
that \textit{for any $J$ the transmission $\clT$ cannot tends toward
zero faster than $L^{-1}$ as $L\rightarrow\infty$.} This theorem
has been proved for the NLSE \eqref{NLSE}, and we prove it here
for the RNLSE \eqref{int-1}, or \eqref{dsa-1} following the way of
Ref. \cite{DeSo86}, modified for the present model \eqref{dsa-1}.
In this sense, this extension of the Devillard-Souillard estimation
of the wave transmission in the framework of
the RNLSE \eqref{int-1} can be considered as a Corollary
of Theorem (3) of Ref. \cite{DeSo86}.
Note also that the present case is simpler,
and the result immediately follows from the Theorem conditions.

We make partition $L=N\Delta x\equiv \sum \Delta $ with constat
randomness $\eta(x_n)=\eta_n$ at each step
$x_n\in\left(n\Delta x\, ,\, n\Delta x+\Delta x\right)$.
Therefore, for every interval $\Delta x_n$ with constant value $\eta_n$
there is a Hamiltonian/energy $H_n$, which produces by the Hamiltonian form
of Eq. \eqref{dsa-1} for each step $n$:
\begin{equation}\label{dsa-4}
H_n=|\partial_x\psi|^2+\omega|\psi|^2+\beta\eta_n|\psi|^4/2\, .
\end{equation}
Since $\psi$ and $\partial_x\psi$ are continuous at edges of
the steps \cite{DeSo86}, we have
\begin{equation}\label{dsa-4-5}
H_{n+1}-H_n=\beta(\eta_n-\eta_{n+1})|\psi|^4/2\, .
\end{equation}
Taking $\Delta\eta=\eta_{\max}-\eta_{\min}$ as
the maximum fluctuation, we obtain from Eq. \eqref{dsa-4-5}
\begin{equation}\label{dsa-5}
|H_{n+1}-H_n|\le\beta\Delta\eta|\psi|^4/2\le
\frac{\beta\Delta\eta}{2\omega^2}H_n^2=AH_n^2\, ,
\end{equation}
where $A=\frac{\beta\Delta\eta}{2\omega^2}$ and
the second inequality is valid for the positive random potential
$\eta(x)\ge 0$.
This inequality yields a decay of the energy $H_n$ with
$n$ not faster than $1/n$  in the limiting
case of $n\gg 1$. We also have from Eq. \eqref{dsa-5}
that the maximum
decay\footnote{From Eq. \eqref{dsa-5}, we have equality
$H_{n+1}-H_n= -AH_{n}^2$.
Denoting $z_n=AH_n$, we obtain
the iteration  $z_{n+1}=z_n(1-z_n)$, which
maps the unit interval $[0\,, 1]$ into itself.
There is a fix point defined by $z^*=z^*(1-z^*)$, and
the iterations converge to $z^*=0$ \cite{Hu82}.
Therefore, the energy $H_n$ decays to zero with the maximal rate
according Eq. \eqref{dsa-6} with the initial condition at the
incident point at $x=0$.}
of the energy reads
\begin{equation}\label{dsa-6}
H_n=\frac{H_0}{1+AH_0n}\, .
\end{equation}
From the partition it follows that $\clT=|\psi_N|^2/|\psi_0|^2$,
and taking into account Eqs. \eqref{dsa-5} and \eqref{dsa-6},
the transmission coefficient reads
\begin{equation}\label{dsa-7}
\clT=\frac{|\psi_N|^2}{|\psi_0|^2}\sim \frac{H_N}{H_0}\ge \frac{1}{1+bL}\, ,
\end{equation}
where $AH_0= b\Delta x$ and $L=N\Delta x$.
Therefore, the transmission $\clT$ cannot tends
toward zero faster than $ L^{-1}$  as  $L$ tends to $\infty$.
This behavior is also supported by numerical investigations
of the scattering problem in Eqs. \eqref{dsa-1} \eqref{dsa-2},
reported in Ref. \cite{SSSES18}.

\section{Fokker-Planck equation}\label{sec:fpe}

In this section, we consider
one-dimensional localization of stationary
solutions of the RNLSE \eqref{int-1} in a random
$\delta$-correlated potential $\eta(x)$
with a Gaussian distribution (white noise), of zero mean
and variance 2D: namely, $\lgl\eta(x)\eta(x')\rgl =2D\delta(x-x')$.
Following Refs. \cite{IoFi07,FiIoMa08},
we study Anderson localization of stationary solutions
with energies $\omega$. In this case, the wave functions are real
$\psi(x)=\psi^*(x)=\phi(x)$.

We will specifically calculate $\lgl\phi^2(x)\rgl$ of solutions of Eq.
\eqref{int-1} that are found for a certain $\omega$,
with given boundary conditions at some point, for example,
$\phi(x=0)$  and $\phi^{\prime}(x=0)$,
where the prime means the derivative with respect to $x$. This
will be done with the help of the analogy with the Langevin
equation \cite{IoFi07,halperin,LiGrPa88}. In particular,
we will calculate the growth rate of the second moments
$\lgl\phi^2(x)\rgl$ and $\lgl(\phi^{\prime}(x))^2\rgl$.
Therefore, considering coordinate $x$ as a formal time on the half axis
with the definition $\sqrt{\omega}x\equiv \tau\in[0\, ,\infty)$,
Eq. \eqref{int-1} reduces to the classical Langevin equation
\begin{equation}\label{fpe-1}
\ddot{\phi}+\phi=\beta_{\omega}\eta(\tau)\phi^3\, , \quad\quad
\beta_{\omega}=\frac{\beta}{\omega}\, .
\end{equation}
Here $\eta(\tau)$ is considered as the $\delta$-correlated
Gaussian noise\footnote{In this notation, the correlator is scaled over
the variance 2D.} $\lgl\eta(\tau)\eta(\tau')\rgl =\delta(\tau-\tau')$.

The dynamical process in the presence of the Gaussian
$\delta$-correlated noise is described by the distribution function
$P=P(u,v,\tau)=\lgl\delta(\phi(\tau)-u)\delta(\dot{\phi}(\tau)-v)\rgl$
that satisfies the Fokker-Planck equation (FPE) \cite{IoFi07}
\begin{equation}\label{fpe-2}
\partial_{\tau}P=-v\partial_uP+u\partial_vP+
\beta^2_{\omega}u^6\partial^2_vP\, .
\end{equation}
The FPE produces an infinite chain of equations for the averages
\[
r_{m,n}=\lgl u^mv^n\rgl=\int dvdu u^mv^nP(u,v,\tau)\, ,
\]
where $m+n=2+4l$ with $l,m,n=0,1,2,\dots$. This
yields a system of equations
\begin{align}\label{fpe-3}
&\dot{r}_{2,0}=2r_{1,1}\, , \nonumber \\
&\dot{r}_{1,1}=-r_{2,0}+r_{0,2}\, , \nonumber \\
&\dot{r}_{0,2}=-2r_{1,1}+2\beta^2_{\omega}r_{6,0}\, , \nonumber \\
&\dot{r}_{6,0}=6r_{5,1}\, . \\
&\vdots \nonumber
\end{align}
In the vector notation, it reads
\begin{equation}\label{fpe-4}
\dot{\bfR}=\hM\bfR\, .
\end{equation}
Here $\hM$ is a dynamical matrix, which will be defined
defined from Eq. \eqref{fpe-3} below in the text.

\subsection{First order of the perturbation theory}

The chain of equations in the system \eqref{fpe-3} is truncated
at the term $r_{6,0}=\lgl u^6\rgl$, taking $\dot{r}_{6,0}=0$ and
$2\beta^2_{\omega}r_{6,0}=h$ is a small constant
value\footnote{In the numerical experiment \cite{SSSES18},
fluctuations of the refractive index $\beta\sim 10^{-3}$.
However it cannot be neglected, since it incorporated
into the highest derivative in the FPE \eqref{fpe-2}.}.

It should be admitted that the rest of this infinite chain
of linear equations, determined by matrix $\hM$, contributes
only to the term $r_{6,0}(\tau)$. However, in the perturbation approach,
when $h=0$, the material is transparent. Therefore, neglecting
$h$-terms in the rest of the matrix $\hM$, we obtain it in the
Jordan block form, where every block matrix $M[(3+4l)\times(3+4l)]$
has the same structure, determined by operator $-v\partial_u+u\partial_v$.
Its integral curves are circles in the $(u,v)$ phase
space \cite{Schutz}. Correspondingly, every block
matrix has imaginary eigenvalues, and
$2\beta^2_{\omega}r_{6,0}\le h=\mathrm{const}$.

Consequently, after the truncation, Eq. \eqref{fpe-4} becomes
inhomogeneous equation with $l=0$, which reads
\begin{equation}\label{fpe-5}
\dot{\bfR}=\hM\bfR+h\bfV\, ,
\end{equation}
where
\begin{equation}\label{fpe-6}
\bfR=\begin{pmatrix} r_{2,0}&\\ r_{1,1}&\\ r_{0,2}&\end{pmatrix}\, ,
\quad
\hM=\begin{bmatrix} {}0& {}2& {}0& \\ -1& {}0& {}1& \\ {}0& -2& {}0&
\end{bmatrix}\, , \quad
\bfV=\begin{pmatrix} 0& \\ 0& \\ 1& \end{pmatrix}\, .
\end{equation}
We take the ``initial'' condition $\bfR_0^{T}=(0,0,1)$. Performing
the Laplace transformation $\tilde{\bfR}(s)=\clL[\bfR(\tau)](s)$, we arrive
at the equation for the Laplace image
\begin{equation}\label{fpe-7}
\tilde{\bfR}(s)=\frac{1}{s(s-\hM)}(\bfR_0s+h\bfV)\, .
\end{equation}
The inverse Laplace transformation yields the solution
\begin{equation}\label{fpe-8}
\bfR(\tau)=e^{\hM\tau}\bfR_0+h\frac{e^{\hM\tau}-1}{\hM}\bfV\, .
\end{equation}
Let us consider the series of the exponential. To this end we use
the Cayley-Hamilton theorem\footnote{According to the
Cayley-Hamilton theorem,
every complex square matrix satisfies its own characteristic equation,
see for example \cite{Lankaster}.}. The characteristic equation of
matrix $\hM$ is $\lambda^3+4\lambda=0$. Therefore, $\hM^3=-4\hM$
and correspondingly $\hM^4=-4\hM^2$.
These equalities determine the exponential in Eq. \eqref{fpe-8}
as follows
\begin{subequations}\label{fpe-9}
\begin{align}
e^{\tau\hM}=1+\frac{1}{4}[1-\frac{1}{4}\cos(2\tau)]\hM^2+
\frac{1}{2}\sin(2\tau)\hM\, ,  \label{fpe-9a}\\
\frac{e^{\tau\hM}-1}{\hM}=\tau +\frac{1}{8}[2\tau-\sin(2\tau)]\hM^2
+\frac{1}{4\tau}[1-\cos(2\tau)]\hM\, . \label{fpe-9b}
\end{align}
\end{subequations}
Applying these expressions to the solution \eqref{fpe-8},
we eventually obtain that $r_{2,0}$ and $r_{0,2}$ increase
linearly for large values of $\tau$. Therefore
the second moments of the wave function and its derivative
grow linearly with the coordinate $x$:
\begin{equation}\label{fpe-10}
\lgl\phi^2(x)\rgl\sim 1+\beta^2_{\omega}\omega^{\frac{1}{2}}x \, .
\end{equation}

Important part of the analysis is the perturbation approach,
which is valid only when the nonlinear randomness/(random nonlinearity)
is small.
The situation becomes completely unclear for the strong nonlinearity.
Therefore. we take into account all the orders
of the perturbation theory.

\subsection{Iteration procedure for higher orders of the
perturbation theory}

Now we consider Eq. \eqref{fpe-5} for the next Jordanian
block with $l=1$, which reads $\dot{\bfR}_1=\hM_1\bfR_1+h\bfV_1$.
It describes moments $r_{m.n}$ with $m+n=6$.
Therefore, $\bfR_1^{T}=(r_{6,0}\, ,r_{6,1}\dots\, , r_{0,6})$
and $V_1^{T}=(0\, ,0\, ,2\, ,6\, ,12\, ,20\, ,30)$.
We are interesting in the dynamics of
$r_{6,0}$, which also grows linearly with $\tau$. Its numerical
solution is shown in Fig.~\ref{fig:fig1}.
\begin{figure}
\centering
\includegraphics[width=8cm]{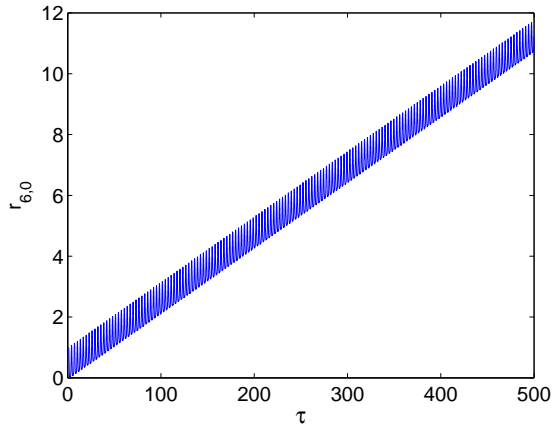}
\caption{(Color online) Dynamics of
the sixth moment $r_{6,0}$ vs time $\tau$ as a
result of numerical solution
of Eq. \eqref{fpe-5} for $l=1$ and $\beta^2_{\omega}=10^{-3}$
(MATLAB, ode45).}
\label{fig:fig1}
\end{figure}
\begin{figure}
\centering
\includegraphics[width=8cm]{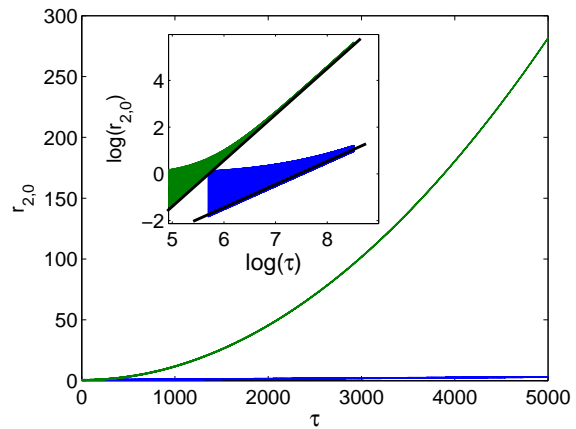}
\caption{(Color online) Dynamics of the second moments
obtained numerically for $\beta^2_{\omega}=10^{-3}$
(MATLAB, ode45). The upper (green) curve
corresponds to the second order of the perturbation
and the lower (blue) curve corresponds to the first
order of the perturbation.}
\label{fig:fig2}
\end{figure}
Therefore, the second order of the perturbation,
shown in Fig.~\ref{fig:fig2},
yields quadratic growth of the moment $r_{2,0}(\tau)\sim \tau^2$.

Since the structures of the Jordanian blocks are just the same for all
$l=0\, ,1\, ,\dots\, ,$, then each $l$-th block has its solution
in the form of Eq. \eqref{fpe-8}, which reads
\begin{equation}\label{fpe-11}
\bfR_l(\tau)=e^{\hM_l\tau}\bfR_0+h\frac{e^{\hM_l\tau}-1}{\hM_l}\bfV_l\, .
\end{equation}
This solution corresponds to the linear growth of the moment
$r_{2+4l,0}(\tau)\sim \beta^2_{\omega}\tau$.
Therefore in the $l$-th order of the perturbation theory,
after $l$ times of integration, we obtain $r_{6,0}(\tau)$
in Eq. \eqref{fpe-5} as follows
\begin{equation}\label{fpe-12}
h\sim \sum_{j=0}^{l}\frac{(\beta^2_{\omega}\tau)^j}{j!}\, ,
\end{equation}
where constants in the primitives at every integration, are taken to be one,
to obtain the truncated series of an exponential function.
Continuing this iteration procedure \textit{ad infinitum},
we eventually obtain the second moment in the exponential
form as follows
\begin{equation}\label{fpe-13}
\lgl\phi^2(x)\rgl=r_{2,0}(\tau)\simeq e^{\beta^2_{\omega}\tau}
=\exp(\beta^2\omega^{-3/2} x)\, .
\end{equation}

\subsection{Anderson localization of the wave function}

The rate of this exponential growth of the second moment
is determined by a so-called generalized Lyapunov
exponent \cite{ZiPi03}
\begin{equation}\label{al-1}
2\gamma = \lim_{x\to\infty}\frac{\ln\lgl\phi^2(x)\rgl}{x}
=\frac{\omega^{3/2}}{\beta^2}\, .
\end{equation}
This exponential growth is a strong indication of exponential,
Anderson, localization of the stationary states $\psi_{\omega}(x)$
with a localization length $\xi=1/\gamma$.
Note that it is different from the usually
studied self-averaging quantity
\begin{equation}\label{al-1a}
\gamma_s=\frac{1}{2}\frac{d}{dx}\lgl\ln\phi^2(x)\rgl=
\frac{1}{2}\lim_{x\to\infty}\frac{\ln\phi^2(x)}{x}\, ,
\end{equation}
which determines a genuine localization
length\footnote{In the linear case,
with a Gaussian noise, these values are related \cite{ZiPi03},
$\gamma=\gamma_s+a$, where $a$  is due to the width of
the Gaussian process.}. In the NLSE \eqref{NLSE} and \eqref{NLSE-a},
the Lyapunov exponent $\gamma$ is independent of $\beta$ and coincides
with the linear limit \cite{IoFi07}. However, here the limit with
$\beta=0$ does not exists and as admitted above, $\beta$
cannot be a parameter for a perturbation consideration.
We will return to this value in Sec.~\ref{sec:npt}, while here
following \cite{IoFi07,FiIoMa08},
we explore \textit{\`a la} Borland arguments: since the
distribution of the random potentials
is translationally invariant, it is independent of the choice of
the initial point as $x=0$. As in the linear case, starting from a
specific initial condition, $\phi(x)$ will typically grow.
For specific values of $\omega$ at some point this function will start to
decay, so that a normalized eigenfunction is found.
Borland's arguments \cite{borland,MoTw61} are rigorous for the
linear case \cite{DeLeSo85}.
In a heuristic form, applied to the
nonlinear case, the envelope of the wave function will grow
exponentially if we start either from the right or from the
left. The value of $\omega$ results from the matching condition, so
that an eigenfunction has some maximum and decays in both
directions as required by the normalization condition.
The
exponential decay is an asymptotic property, while the
matching is determined by the potential in the vicinity of the
maximum. This observation is crucial for the validity of this
approach and enables us to determine the exponential decay
rate of states from the solution of the initial value problem
in the form of both NLSE \eqref{NLSE} and RNLSE \eqref{int-1}.

\section{Numerical perturbation theory}\label{sec:npt}

In this section we develop an analytical approach based
on the numerical evaluation of a small parameter, which
is necessary for the perturbation theory. We rewrite Eq.
\eqref{int-1} in the form
\begin{equation}\label{LA-1}
-\hh^2\partial_x^2\psi +\beta\eta(x)|\psi|^2\psi=\omega\psi\, ,
\end{equation}
where we introduced a dimensionless Planck constant $\hh$, which should be
small enough for correct numerical evaluation of Anderson localization.

Parameter $\beta$ cannot be zero: $\beta\neq 0$ since
it does not fulfill the boundary condition $\psi(x=\pm\infty)=0$
for the equation $\partial_x^2\psi+\omega\psi=0$, if $\beta=0$.
Therefore a perturbation theory over $\beta$ does not exists
even for small $\beta\ll 1$. Therefore we shall keep $\beta=1$.

Let us perform an identity transformation by adding and subtracting
the linear random term $\beta\eta(x)\psi(x)$ in Eq. \eqref{LA-1}.
We have
\begin{multline}\label{LA-2}
\hh^2\partial_x^2\psi+\omega\psi-\beta\eta(x)\psi(x) \equiv \\
\equiv\omega\psi - \hat{H}_0\psi
 =-\beta\eta(x)\psi(x)+\beta\eta(x)\psi^3\, ,
\end{multline}
where $\hat{H}_0$ is the Anderson Hamiltonian. Therefore,
the l.h.s. of Eq. \eqref{LA-2} corresponds to the
eigenvalue problem of Anderson localization
\begin{equation}\label{LA-3}
\hat{H}_0\varphi_n(x)=[-\hh^2\partial_x^2+\beta\eta(x)]\varphi_n(x)=
\sigma_n\varphi_n(x)  \, ,
\end{equation}
where the eigenvalues are functions of $\beta$,
namely $\sigma_n=\sigma_n(\beta)$ and
\begin{equation}\label{LA-4}
\int_{-\infty}^{\infty}\varphi_m(x)\varphi_n(x)dx=\delta_{m,n}\, .
\end{equation}

The Hamiltonian $\hat{H}_0$ is Hermitian, and
$\{\varphi_n(x)\}$ is a complete set of
orthogonal functions.
Therefore, the stationary states can be expanded over the Anderson modes
\begin{equation}\label{LA-5}
\psi(x)\equiv\psi_{\omega}(x)=\sum_na_n(\omega)\varphi_n(x)\, .
\end{equation}

Substituting expansion \eqref{LA-5} in Eq. \eqref{LA-2}, we obtain
\begin{multline}\label{LA-6}
\omega\sum_na_n\varphi_n(x)-\sum_n\sigma_n\varphi_n(x) =
-\beta\eta(x)\sum_na_n\varphi_n(x)- \\
-\beta\sum_{n_1,n_2,n_3}
a_{n_1}a_{n_2}a_{n_3}\varphi_{n_1}(x)\varphi_{n_2}(x)\varphi_{n_3}(x)\, ,
\end{multline}
where $a_n\equiv a_n(\omega)$.

\subsection{Algebraic equation and small parameter}\label{AESP}

Multiplying Eq. \eqref{LA-6} by
$\varphi_m(x)$ and integrating with respect to $x$, and taking into
account Eq. \eqref{LA-4}, we arrive at the algebraic equation
\begin{multline}\label{AESP-1}
a_m(\omega-\sigma_m) =
-\beta\sum_nA_{m,n}a_n- \\
+\beta\sum_{n_1,n_2,n_3}B^m_{n_1,n_2,n_3}a_{n_1}a_{n_2}a_{n_3}\, ,
\end{multline}
where the overlapping integrals are
\begin{subequations}\label{AESP-2}
\begin{align}
& A_{m,n}=\int\eta(x)\varphi_m(x)\varphi_n(x)dx\, ,
\label{AESP-2a} \\
&B^m_{n_1,n_2,n_3}= \int\eta(x)
\varphi_{n_1}(x)\varphi_m(x)\varphi_{n_2}(x)\varphi_{n_3}(x)dx\, .
\label{AESP-2b}
\end{align}\end{subequations}

The overlapping integrals in Eqs. \eqref{AESP-2} are estimated
numerically, and the results are presented in
Figs.~\ref{fig:fig3} and \ref{fig:fig4}.
These integrals are definitely less than $1$, and we take
them as the first order of the approximation by small
parameter\footnote{Some heuristic arguments for supporting
this statement are on turn.
If the localization length is large, the segment of integration
is large enough to apply the ergodic theorem for the Markov process,
which yields $\int \eta(x)dx\approx \langle \eta(x)\rangle =0$.
Therefore, the integrals can be considered as a small parameters $\vep$.
Moreover, the larger segment of integration is, the smaller $\vep$ is,
and we can account only diagonal and few nearest of-diagonals in the
matrices $A$ and $B$.
In the opposite case, when the localization length is small, then
overlapping of Anderson modes is small, and only diagonal and
nearest neighbors matrix elements should be taken into account.
The off-diagonal elements are small values $\vep\ll 1$, however for
the diagonal elements the integrals are just less than $1$ and
these terms contribute to the energy $\omega$. Integrals $B^m_{n_1,n_2,n_3}$
are four-tensors, however in the first order of $\vep$, only terms with $n_1=n_2=n_3=n$ are accounted.} $\vep$:
$\vep<A_{m,n}\sim B^m_{n_1,n_2,n_3}< 1$.

\begin{figure}
\centering
\includegraphics[width=8cm]{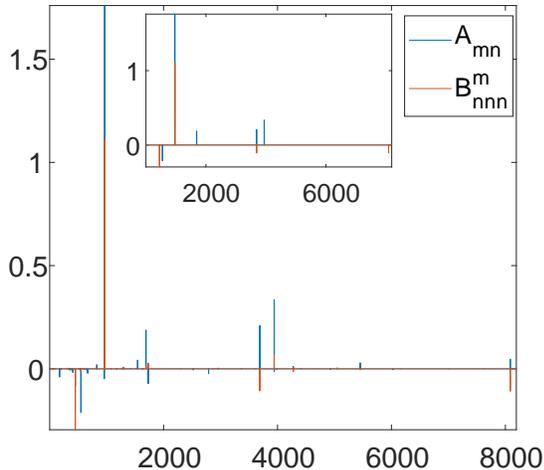}
\caption{(Color online) Overlapping integrals $A_{m,n}$ and $B^m_{n,n,n}$
for the random mode $m=965$ with $\sigma_m=-2.8822$.
The Anderson modes are result of numerical
calculation of Eq. \eqref{LA-3} on a chain with zero
boundary condition, number of cites $N=8192$, $\hh=1.5915\cdot 10^{-8}$.
Insert presents overlapping integrals, which contribute to the spectrum
for $n=m$ and stationary states $\psi_{\omega}(l)$ for $n\neq m$ in the first order of perturbation, when the absolute values of the overlapping integrals
are larger then $\vep=0.1$.  }
\label{fig:fig3}
\end{figure}

\begin{figure}
\centering
\includegraphics[width=8cm]{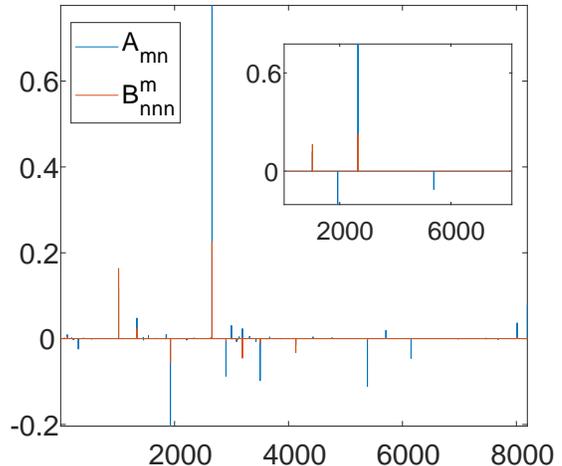}
\caption{(Color online) The same as Fig.~\ref{fig:fig3}
for the random mode $m=2657$ with $\sigma_m=-1.6727$.}
\label{fig:fig4}
\end{figure}

\subsection{Perturbation theory}

After taking into account only nearest neighbors of the $m$-th mode,
algebraic equation \eqref{AESP-1} reads
\begin{multline}\label{AESP-3}
 a_m(\omega-\sigma_m) =
 -\beta A_{m,m}a_m+\beta B^m_{m,m,m}a^3_m \\
 -{\sum_n}^{\prime}[\beta A_{m,n}a_{n}-\beta B^m_{n,n,n}a^3_n]\, .
\end{multline}
where prime means $n\neq m$.
Taking into account that the overlapping integrals in Eqs.
\eqref{AESP-2} are small, we cast
solution for $a_n$ in form of the
decomposition/expansion over $\vep$ up to the first order of $\vep$:
\begin{equation}\label{AESP-5}
a_m=a_m^{(0)}+\vep a_m^{(1)}\, \quad\quad m=1\, , \dots\,, N \, .
\end{equation}
Substituting Eq. \eqref{AESP-5} in Eq. \eqref{AESP-3} and collecting
terms with corresponding orders of $\vep$,
we have for the zero order
\begin{equation}\label{AESP-6}
a_m^{(0)}(\omega-\sigma_m)=0 \, ,
\end{equation}
which yields $a_m^{(0)}(\omega)=1$ for $\omega=\sigma_m$ and
$a_m^{(0)}(\omega)=0$ for $\omega\neq\sigma_m$
and $m=1\, , \dots\,, N$. Formally, we define it as a Kronecker
delta:
\begin{equation}\label{AESP-7}
a_m^{(0)}\equiv a_m^{(0)}(\omega)=\delta(\omega-\sigma_m),
\quad \quad m=1\, ,2\, ,\dots\, , N \, .
\end{equation}
For the first order of $\vep$, we have
\begin{equation}\label{AESP-8}
a^{(1)}_m(\omega-\sigma_m)=
-\beta{\sum_n}^{\prime}[A_{m,n}a^{(0)}_{n}-
\beta B^m_{n,n,n}(a^{(0)}_n)^3] \, ,
\end{equation}
This equation has solution only for $\omega=\sigma_{n}$
with $n\neq m$ that yields
\begin{equation}\label{AESP-9}
a^{(1)}_m\left(\omega=\sigma_{n}\right)
=\frac{\beta}{\sigma_m-\sigma_{n}}\, ,
\end{equation}
while the spectrum reads
$\omega=\omega_n=\sigma_m-\beta (A_{m,m}-B^m_{m,m,m})$.
Here we used that $a^{(0)}_m(\omega=\sigma_{n})=0$.

Eventually, we obtain that the stationary states
$\psi_{\omega}(x)$
are localized only for the discrete spectrum
$\omega\in {\rm spec}
\left(\sigma_n+B^{m}_{n,n,n}-A_{m,n}\right), ~n=1,2,\dots$.
For example, for $\omega=\sigma_n$, the stationary
state $\psi_{\sigma_n}(x)$ contains all $a_{n}$-s with
$\omega=\omega_n$, which
are $a_n^{(0)}(\sigma_n)$ and $a_{n}^{(1)}(\sigma_n')$.
Therefore, the stationary state reads
\begin{equation}\label{AESP-10}
\psi_{\omega_n}(x)=\varphi_{\sigma_n}(x)
+\beta\sum_{k\neq n}\frac{\varphi_{\sigma_k}(x)}{\sigma_{n}-\sigma_{k}}
\left(A_{n,k}-B^n_{k,k,k}\right)\, ,
\end{equation}
with eigenenergy
\begin{equation}
\omega_n=\sigma_n-\beta (A_{n,n}-B^n_{n,n,n}) \, .
\end{equation}
Results of numerical evaluation of two random
stationary modes
are presented in Figs.~\ref{fig:fig5} and \ref{fig:fig6}.

\begin{figure}
\centering
\includegraphics[width=8cm]{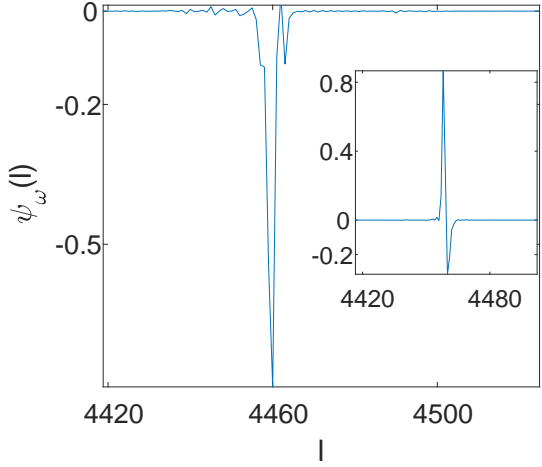}
\caption{(Color online) The normalized stationary state $\psi_{\omega_m}(l)$
constructed near the Anderson mode (in insert) $\varphi_m(l)$
for $m=965$ with $\sigma_m=-2.8822$.}
\label{fig:fig5}
\end{figure}

\begin{figure}
\centering
\includegraphics[width=8cm]{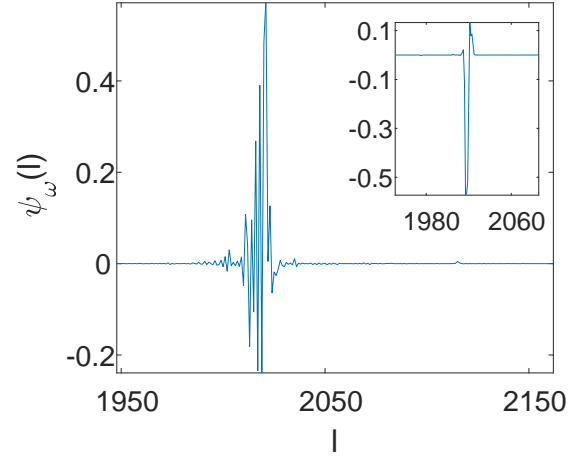}
\caption{(Color online) The same as Fig.~\ref{fig:fig5}
for $m=2657$ and $\sigma_m=-1.6727$.}
\label{fig:fig6}
\end{figure}

\section{Conclusion}\label{sec:cocl}

We considered two related to each other problems, and
obtained two results. The first one is the power law decay
of the transmission coefficient as a function of the slab length $L$.
This result is obtained in both heuristic and microscopic (in the
framework of the Helmholtz-RNLSE) approaches. We present these alternative
ways to explain the experimental setup \cite{SSSES18} of
wave propagation in nonlinear media.
The second result corresponds to the general microscopic
consideration of Anderson localization in the
RNLSE \eqref{int-1}. We constructed
a perturbation theory over the nonlinearity parameter.
Constructing an iteration procedure, we were able
to take into account all orders of the perturbation
theory by means of a resummation procedure over all the orders
of the small parameter $\beta^2_{\omega}$.
As the result of the developed method,
we obtained exponential (Anderson) localization
of the wave function $\phi(x)\sim e^{-x/\xi}$,
estimating the mean squared second moment growth.
It should be stressed that this approach does not determined the
localization length $\xi$, and we estimate it approximately
according to the generalized
Lyapunov exponent $\gamma$: $\xi=1/\gamma
\sim 1/2\sqrt{\omega}\beta^2_{\omega}\gg 1$\footnote{We stress that
it is not a genuine localization length, defined
above in Eq. \eqref{al-1a}.}.

We also suggested numerical verification of Anderson
localization of the stationary states $\psi_{\omega}(x)$ for discrete
spectrum $\omega$. To this end we introduced a
linear counterpart of the RNLSE in the form of Eqs. \eqref{LA-2}
and \eqref{LA-3}.  In this case a small parameter
$\vep$ can be introduced in the form of overlapping
integrals of the linear Anderson modes.
This makes it possible to construct a perturbation theory,
which is well controlled numerically\footnote{It is
a first step of the numerical approach. For example,
numerical study of the localization length
as a function of the spectrum $\xi=\xi(\omega)$
is an important task, which should be considered separately.}.

In conclusion, we admit that the obtained results
on the power law and exponential decays correspond to
two different approaches, which are so called "fixed
output" and "fixed input" [7].
The former, considered in Sec. III, leads to the power
law localization, while the latter, considered in
Secs. IV and V, leads to the exponential (Anderson)
localization. Therefore,
there is an essential difference between Eq. \eqref{dsa-1} and
Eq. \eqref{fpe-1}. Namely, in the former case there is a finite
constant probability current that corresponds to the fixed
output conditions, while in the latter case,
the probability current is zero
that is a necessary condition of localization.


This paper is based on our previous work and numerous discussions
with Prof. Shmuel Fishman,
to whose memory I express my deep gratitude and respect.
I am thankful to Mordechai Segev for reading the paper
with helpful and valuable commenting it and to
Alexander Dikopoltzev, Yaakov Lumer for
helpful discussions.
I also thank Yonatan Sharabi for
illuminating and helpful discussions and help in numerical
calculations in Sec.~\ref{sec:npt}.

\end{document}